\documentclass[a4paper,11pt]{article}
\pdfoutput=1 

\usepackage{jinstpub} 
\def\DAFNE{DA$\Phi$NE}

\title{\boldmath The reserach activity of the Frascati Laboratory}

\author[a]{P. Gianotti}

\affiliation[a]{Laboratori Nazionali di Frascati of INFN,\\ via E. Fermi 54, 00044 Frascati, Italy}

\emailAdd{paola.gianotti@lnf.infn.it}

\abstract{
The Frascati National Laboratory (LNF) is the largest and the oldest among the National Laboratories of the Italian Institute for Nuclear Physics (INFN). Since its foundation in 1954, it has been devoted to two main 
activities: the development, construction and operation of particle accelerators; the design and construction of forefront detectors for particle, nuclear and astroparticle physics. The research program of LNF is focused on fundamental research, but interdisciplinary activity has grown of importance along the years, with a perfect balance between internal and external activities. 
The scientific program taking place at LNF, at present, is still centered on the \DAFNE~complex, but in the last years, a second accelerator infrastructure, SPARC\_LAB, devoted to the study and development of new technique of particle acceleration is marking the path toward the future: EuPARXIA. This will be an European infrastructure for plasma acceleration development. 
In this paper, an overview of the research program of the Laboratory and of the future perspectives is presented.
}


\arxivnumber{} 

\proceeding{International Conference "Instrumentation for Colliding Beam Physics" (INSTR-20) \\
  24-28 February 2020\\
  Budker Institute of Nuclear Physics, Novosibirsk, Russia}

\begin{document}
\maketitle
\flushbottom

\section{Introduction}
\label{sec:intro}
The Frascati National Laboratory (LNF) is the largest and the oldest among the INFN National Laboratories. It has been founded in 1954 with the idea of furthering particle physics research in Italy by constructing the first national accelerator: a 1.1 GeV electron synchrotron. LNF is also known worldwide since it is where the first electron-positron collider, named AdA, has been built. Thanks to this successful pioneering initiative, the particle physics community moved into the era of colliders that has seen a long list of machines of increasing energy and complexity culminated with the CERN LHC. At LNF after AdA, other electron-positron colliders were in operation for users: ADONE (1969-1993) and then \DAFNE~(2002-now).

Since the early days, together with the development, construction and operation of accelerators, LNF activity has been strongly committed to the design and construction of forefront detectors for particle, nuclear and astroparticle physics.

\begin{figure}[h]
\centering 
\includegraphics[width=0.8\textwidth,clip]{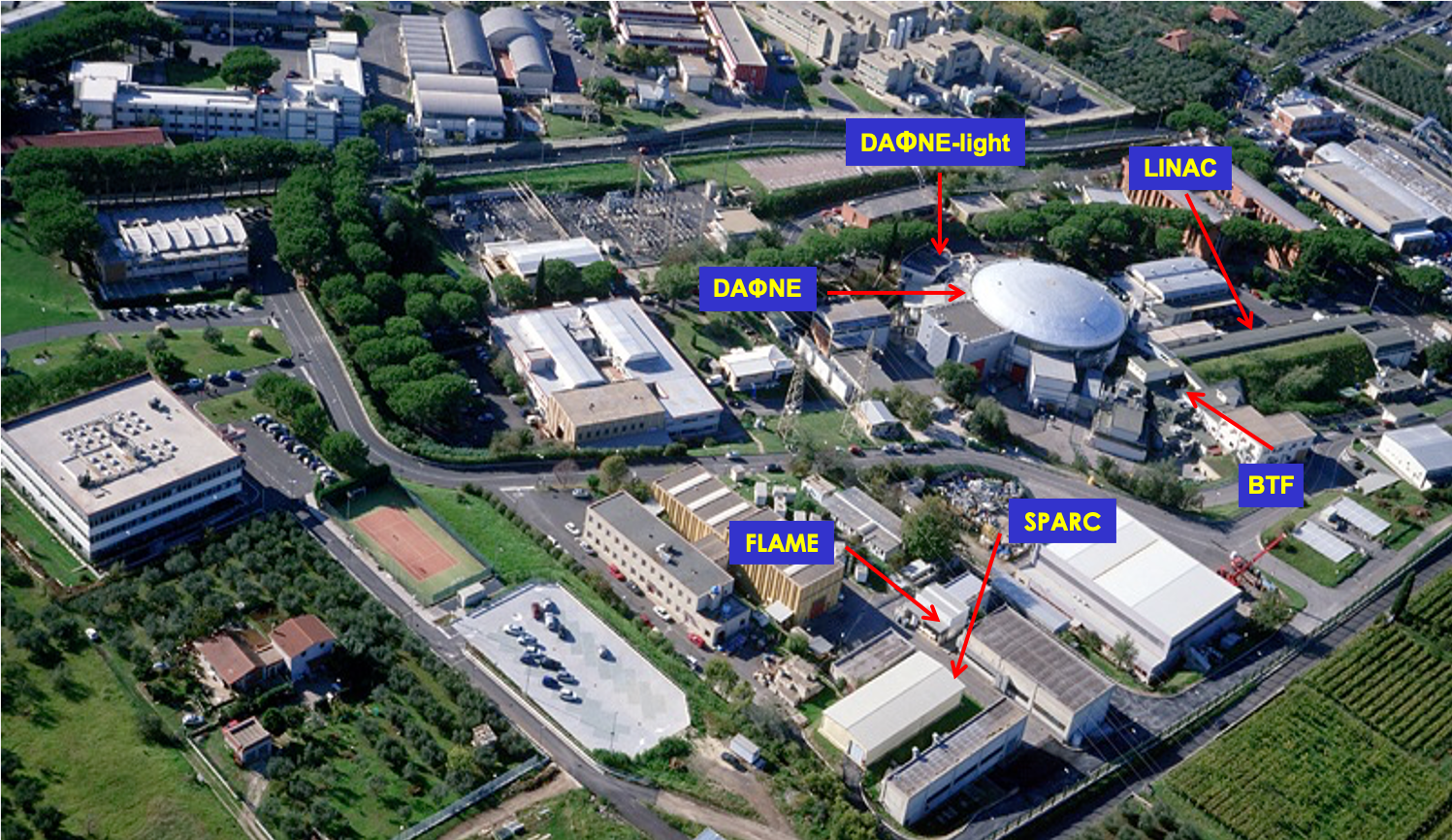}
\caption{\label{fig:1} LNF aerial view; the main infrastructures of the laboratory are pointed out (see text for more details).}
\end{figure}

Today LNF stands over an area of 135.000 mq of which 56.000 are indoor and includes offices, laboratories and service areas.  Figure~\ref{fig:1} shows an aerial view of LNF premises with the main infrastructures pointed out. They will be described in the following sections together with the present scientific program.

\section{Infrastructures and scientific program}
The central focus of the activities carried out at LNF have always been in the fields of high energy physics, accelerator operation and development, and interdisciplinary research, with a perfect balance between internal activities, carried out on site, and external ones taking place in the major laboratories all over the world.

The LNF site hosts several research infrastructures:
\begin{itemize}

\item \DAFNE, an $e^+ e^-$ collider, unique in Europe, operated at the $\Phi$ energy and able to deliver instantaneous luminosities $\mathcal{L} \sim 2 \times 10^{32}$ cm$^{-2}$ s$^{-1}$;
\item  \DAFNE-light, a synchrotron light laboratory, housing several synchrotron radiation lines extracted from the electron ring of \DAFNE~in the soft-X and infrared range;
\item a Beam Test Facility, BTF, an experimental area equipped for detector and beam diagnostic tests, where two beam lines are available which can provide electrons, positrons and photons of variable intensity and energy;
\item SPARC\_LAB, a complex hosting a linear electron accelerator (SPARC) able to drive a FEL, and a laser (FLAME) of power $\sim$ 200 TW. The SPARC\_LAB is an infrastructure for R\&D on new techniques of particle acceleration and for interdisciplinary studies, including PWFA and LWFA experiments, TeraHertz radiation and a Compton source;
\item SCF\_LAB, a certified laboratory for the design, modelling and characterization of laser ranging equipment. The laboratory procedures are approved by the International Laser Ranging Service;
\item COLD - CryOgenic Laboratory for Detectors, an infrastructure for R\&D of ultra-sensitive photon detectors in the range 10-100 GHz, mainly devoted to the search for Axions. Novel superconducting resonant cavities operating at 10-20 GHz at one side, and development of Josephson Junction devices for 50-100 GHz microwave photons on the other side, are the topics presently under study, with the aim of reaching single photon detection capability;
\item a laboratory for the development of gaseous detectors of novel concept - Detector Development Group, DDG. This infrastructure has its roots in the long standing tradition of LNF in the R\&D, design and manufacturing of gaseous detectors, such as wire tubes operated in proportional or streamer mode (1985-1990), RPC with glass electrodes (1991-1994), large drift chamber (1995-1997) and Micro-Pattern-Gaseous-Detector (MPGDs - since 2000) for large high-energy-physics experiments~\cite{a0};
\item large assembly halls with several clean rooms (for a total surface of about 480 mq) equipped with special tools for designing and building large experimental equipment. These have been used in the recent past, to build the LAV detector for the CERN experiment NA62~\cite{a1}, and parts of the LHC apparata ATLAS, LHCb and ALICE. At present, several of the detector upgrades foreseen for LHC phase2 are taking place in these premises.
\end{itemize}
Furthermore, LNF hosts a mechanical workshop, an electronic service, a powerful and modern computing center, involved in the GRID Tier2 project, and a medical physics service unique within INFN.

In the following sections, more details will be given on the \DAFNE~accelerator complex, that represents the heart of the present on-site activities of LNF, and on the SPARC\_LAB that on the other hand is emerging as a promising facility to maintain LNF at the forefront of international fundamental research.

\section{The \DAFNE~complex}
\begin{figure}[h]
\centering 
\includegraphics[width=0.8\textwidth,clip]{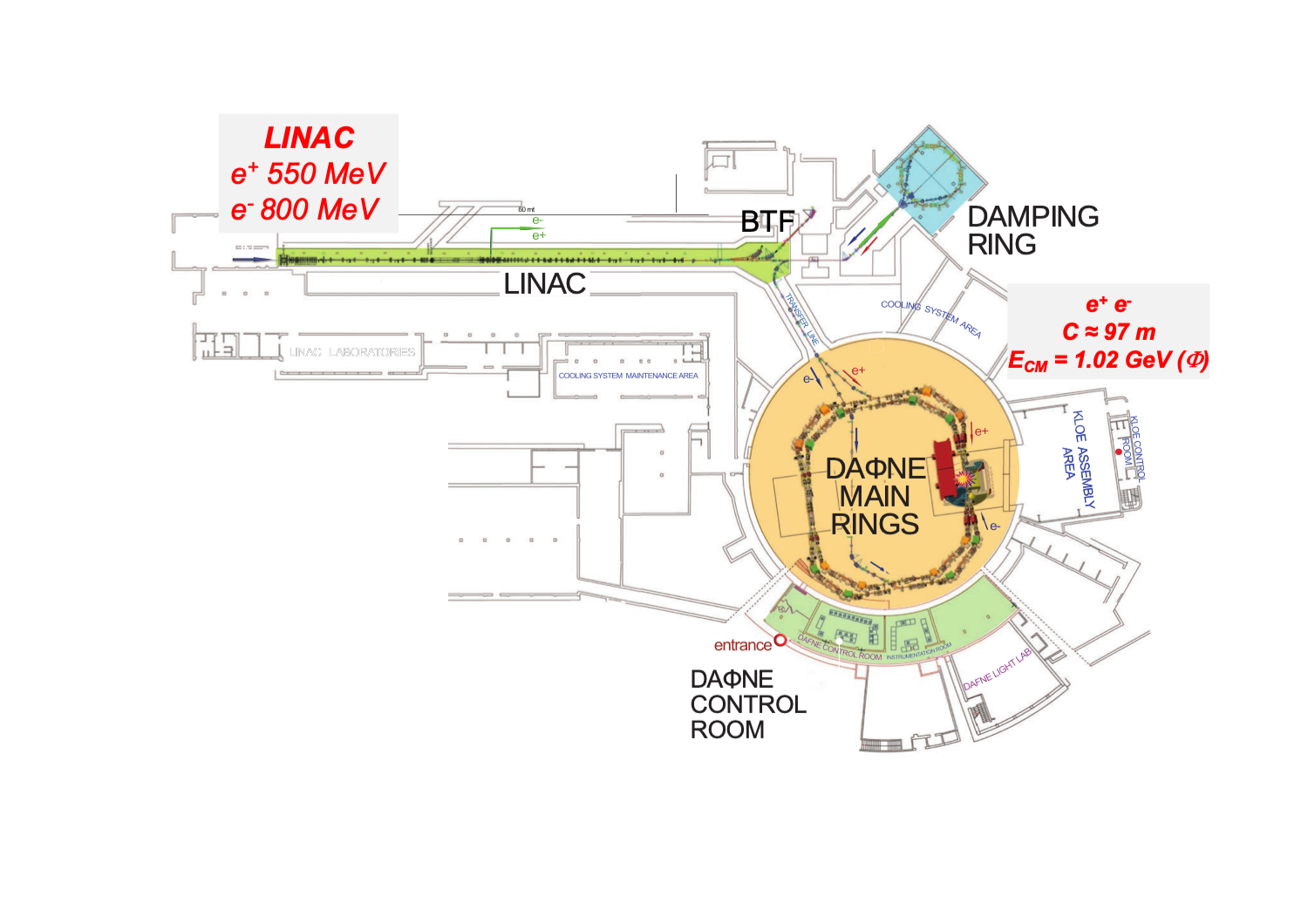}
\caption{\label{fig:2} Layout of the \DAFNE~ complex after the upgrade of 2013 (see text for more details).}
\end{figure}

\DAFNE~ (Double Annular $\Phi$ Factory for Nice Experiments)~\cite{a} is a $\Phi$-factory, {\it i.e.} an $e^+ e^-$ collider with a fixed c.m. energy of 1.02 GeV in order to excite the $\Phi$ meson. The main objective of this accelerator is to provide high rates of neutral and charged kaons to allow dedicated studies of $CP$-violation (KLOE/KLOE-2 experiments~\cite{aa}) and of low-energy interactions of charged kaons with nuclear matter (FINUDA and SIDDHARTA experiments~\cite{aaa}). This is possible since the $\Phi$ resonance decays with a branching fraction $\sim 50\%$ into a couple of charged $K$ mesons, and $\sim 35\%$ in a couple of neutral $K_{long} + K_{short}$.
Furthermore, the \DAFNE~ complex includes a S-band linac, 180 m long, and an accumulator/damping ring,  allowing a fast  and efficient top-up electron/positron injection in the main rings. Figure ~\ref{fig:2} shows the layout of the \DAFNE~ complex.

\DAFNE~started operation in 1999 and since then different experiments alternated on the two opposite Interaction Points (IP1 and IP2). In 2013, \DAFNE~ undertook a general consolidation program aimed at improving its luminosity and the  machine layout was modified  to allow interactions only at IP2. A new Crab-Waist (CW) collision scheme~\cite{b}  was implemented for the first time ever. Thanks to this, a peak luminosity $\mathcal{L} = 4.5 \times 10^{32}$cm$^{-2}$s$^{-1}$ was reached, two orders of magnitude higher than the one measured by other colliders working at the same c.m. energy.

At present, the apparatus installed on \DAFNE~ IP2 is SIDDHARTA-2 (Silicon Drift Detector for Hadronic Atom Research by Timing Application). This experiment aims at performing precise measurements of antikaon-nucleon scattering lengths~\cite{c}.

An other important element of the \DAFNE~ complex is the BTF~\cite{d}. This is a transfer line optimized for selection, attenuation and manipulation of electrons and positrons extracted from the linac. 
These beams are transported to the BTF experimental halls (BTFH1, BTFH2) mainly for test-beam purposes, but since 2018, BTFH1 has been exclusively dedicated to the experiment PADME (Positron Annihilation into Dark Matter Experiment)~\cite{e}. 
The main goal of PADME is to search for dark photons of mass up to 23.7 Mev$/c^2$ produced in the annihilation of the BTF positron beam with the electrons at rest of a thin active diamond target. The PADME data taking is expected to last for two years.

On average, the current circulating in the \DAFNE~electron ring exceeds 1~Ampere and produces an intense emission of photons ranging from infrared to soft X-rays. Thanks to  this,  \DAFNE~ is  also  used  in  dedicated  or  in  parasitic mode,  for  Synchrotron Radiation  (SR) applications.  In  the  \DAFNE-Light  SR  facility  there  are  several beam lines available for users.  Two  of  them,  the  soft  X-ray  (DXR-1)  and  the  UV  one (DXR-2) extract the radiation form the  \DAFNE~wiggler magnets, while the other beam lines, working in the infrared (IR) region, collects the radiation from a bending dipole. 

The \DAFNE-Light  laboratory is inserted in an international network of SR facilities~\cite{f} and tens of users come every year at LNF to exploit the unique opportunities offered here to many research disciplines. All the available beam-lines are equipped with world-class instruments: microscopes, interferometers, crystal-monochromators, that can also be used with conventional sources when the beam is not circulating in \DAFNE. 

A special mention deserves the High Energy Beamline (HEB) that recently has been subject of a MoU with CERN. The aim of this parntership is the realization of WINDY (White lIgth liNe for Desorption Yields) a dedicated branch-line to allow the study of the effects of the interactions between synchrotron radiation and technical surfaces. 

\section{SPARC\_LAB}
SPARC\_LAB (Sources for Plasma Accelerators and Radiation Compton with Laser And Beam) is the second accelerator facility of LNF. It is based on a conventional high brightness RF photoinjector, SPARC, combined with a multi-hundred terawatt laser, FLAME, and it is devoted to the study of new techniques for the realization of compact particle accelerators.

The SPARC photo-injector can produce and accelerate high-brightness electron beams up to 170 MeV, which feed a 12 m long undulator for FEL generation. Thanks to the versatility of the photo-injector and exploiting the RF based compression technique (velocity bunching), a tunable and intense THz source is also available at SPARC\_LAB for advanced longitudinal beam diagnostics and user experiments. 

In the last years, studies and experiments on plasma acceleration have become the main activity in this infrastructure, thanks also to the availability of  the laser FLAME which can produce pulses carrying 5 J of energy, compressed down to duration of 25 fs. The ultra-intense laser pulses are employed to study the interaction with matter for many purposes: electron acceleration through LWFA, ion and proton generation exploiting the TNSA mechanism, study of new radiation sources and development of new electron diagnostics, and can work combined with the linac, {\it i.e.} in a Thomson back-scattering experiment, to generate a quasicoherent, monochromatic X-ray radiation~\cite{g}. 

The SPARC\_LAB is involved in many international projects, in particular CompactLight~\cite{h} and EuPRAXIA~\cite{i} for which  LNF has put forward its candidature to become the hosting site for the machine configuration resulting from the EuPRAXIA Design Study~\cite{ii}. 
Within this scenario, LNF is moving into the direction of making this project the principal on-site activity for the future and to be able to reach this ambitious objective, it started the EuPRAXIA@SPARC\_LAB project~\cite{l}. 
In this context, some important preparatory actions are underway at LNF:
\begin{itemize}
\item    realization of a new infrastructure with the size of about 135 $\times$ 35 mq, as the one required to host the EuPRAXIA facility;
\item    design and construction of the first-ever 1 GeV X-band RF linac and of an upgraded FLAME laser up to 0.5 PW;
\item    design and construction of a compact FEL source, equipped with a user beam line at 4-2 nm wavelength, driven by a high gradient plasma accelerator.
\end{itemize}

\begin{figure}[h]
\centering 
\includegraphics[width=0.8\textwidth,clip]{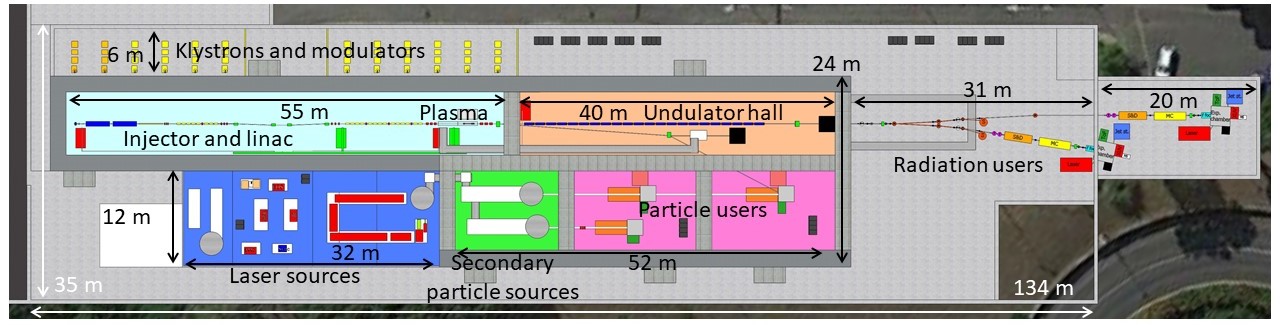}
\caption{\label{fig:3} Layout of the EuPRAXIA@SPARC\_LAB infrastructure.}
\end{figure}

The EuPRAXIA@SPARC\_LAB facility (fig.~\ref{fig:3}) by itself will equip LNF with a unique combination of a high brightness GeV-range electron beam generated in a state-of-the-art linac, and a 0.5 PW-class laser system. Even in the case of LNF not being selected and/or of a failure of plasma acceleration technology, the infrastructure will be of top-class quality, user-oriented and at the forefront of new technologies for particle accelerators. 
EuPRAXIA@SPARC\_LAB is in fact conceived by itself as an innovative and evolutionary tool for multidisciplinary investigations in a wide field of scientific, technological and industrial applications.

\acknowledgments
I would like to thank all the LNF colleagues that helped me preparing this overview.



\begin{thebibliography}{99}
\bibitem{a0}
M. Giovannetti {\it et al.}, \emph{The $\mu$-RWELL for high rate applications}, This proceedings.
\bibitem{a1} P.Massarotti {\it et al.}, \emph{The Large-Angle Photon Veto System for the NA62 Experiment at CERN}, \emph{PoS}  {\bf ICHEP2012} (2013) 504.
\bibitem{a}
G. Vignola {\it et al.}, \emph{\DAFNE, the first $\Phi$-factory}, \emph{Conf. Proc.} {\bf C 960610} (1996) 22-26.
\bibitem{aa} 
F. Curciarello {\it et al.}, \emph{The KLOE-2 e+e- tagging for two-photon physics}, This proceedings.
\bibitem{aaa}
V. Lucherini ({\it for the FINUDA collaboration}), \emph{The FINUDA experiment on DAFNE}, \emph{Int. J. Mod. Phys.} {\bf A 26} (2011) 420-425.\\
J. Zmeskal ({\it for the SIDDHARTA collaboration}), \emph{Kaonic atoms at DAFNE: The SIDDHARTA experiment}, \emph{Int. J. Mod. Phys.} {\bf A 24} (2009) 190-197.
\bibitem{b}
P. Raimondi  {\it et  al.},  \emph{Beam-Beam Issues for Colliding Schemes with Large Piwinski Angle and Crabbed Waist}, arXiV:physics/0702033;  \\
C. Milardi  {\it et al.}, \emph{Present status of the \DAFNE~ upgrade and perspectives}, \emph{Int. J. Mod. Phys.} {\bf A24} (2009) 360-368;\\
M. Zobov {\it et al.}, \emph{Test of ``Crab-Waist'' Collisions at the \DAFNE~ $\Phi$ Factory}, \emph{Phys. Rev. Lett.} {\bf 104} (2010) 174801. 
\bibitem{c}
C. Curceanu {\it et al.}, \emph{Kaonic Atoms to Investigate Global Symmetry Breaking}, \emph{Symmetry} {\bf Vol. 12 Iss. 4} (2020)  547.
\bibitem{d}
B. Buonomo {\it et al.}, \emph{The Frascati Linac Beam-Test Facility (BTF) performance and upgrades}, \emph{Proceedings of the 5th IBIC conference}, 11-15 September 2016, Barcelona, Spain.
\bibitem{e}
P. Gianotti ({\it For the PADME collaboration}), \emph{The investigation on the dark sector at the PADME experiment}, \emph{Nucl. Instrum. Meth.} {\bf A 936} (2019) 266-267.\\
D. Domenici ({\it For the PADME collaboration}), \emph{The PADME detector at LNF}, This proceedings.
\bibitem{f}
\emph{\url{https://www.wayforlight.eu/en/}}
\bibitem{g}
R. Pompili {\it et al.}, \emph{Recent results at SPARC\_LAB}, \emph{Nucl. Instrum. Meth.} {\bf A 909} (2018) 139-144.
\bibitem{h}
\emph{\url{https://www.compactlight.eu/Main/HomePage}}
\bibitem{i}
\emph{\url{http://www.eupraxia-project.eu/}}
\bibitem{ii}
\emph{\url{http://www.eupraxia-project.eu/eupraxia-conceptual-design-report.html}}
\bibitem{l}
D. Alesini {\it et al.}, \emph{EuPRAXIA@SPARC\_LAB Conceptual Design Report}, INFN-18-03/LNF (2018).





\end{thebibliography}
\end{document}